\documentstyle[multicol,aps,epsf]{revtex}

\begin{document}
\title{Slow Coarsening in an Ising Chain with Competing Interactions}
\author{S.~Redner and P.~L.~Krapivsky}

\address{Center for Polymer Studies and Department of Physics, Boston
University, Boston, MA 02215, USA}
\maketitle
\begin{abstract}
  
  We investigate the zero-temperature coarsening dynamics of a chain of Ising
  spins with a nearest-neighbor ferromagnetic and an $n^{\rm th}$-neighbor
  antiferromagnetic interactions.  For sufficiently large antiferromagnetic
  interaction, the ground state consists of $n$ consecutive up spins followed
  by $n$ down spins, {\it etc}.  We show that the asymptotic coarsening into
  this ground state is governed by a multispecies reactive gas of elementary
  excitations.  The basic elementary excitations are identified and each
  decays at a different power-law rate in time.  The dominant excitations are
  domains of $n+1$ spins which diffuse freely and disappear through processes
  which are effectively governed by $(n+1)$-particle annihilation.  This
  implies that the ground state is approached slowly with time, as
  $t^{-1/n}$.

\bigskip 
{PACS number(s): 05.50.+q, 75.10.Hk, 82.20.Mj}
\end{abstract}
\begin{multicols}{2}

\section{INTRODUCTION}

Ising models with nearest-neighbor ferromagnetic and more distant
antiferromagnetic interactions exhibit rich magnetic
ordering\cite{elliott,fisher,bak}.  The competition between ferromagnetism
and longer-range antiferromagnetism leads to different ordered states and an
associated sequence of phase transitions as a function of these two
interaction strengths.  Such models were originally formulated to help
describe the complex magnetism of the rare earths\cite{rare}.  Their unusual
magnetic ordering is believed to arise from the RKKY interaction\cite{rkky},
in which the exchange interaction between localized magnetic moments
oscillates between ferromagnetic and antiferromagnetic as a function of their
separation.  Some of the essential consequences of this complex situation
seem to be captured by Ising models with competing interactions.
  
One of the simplest versions is the axial next-nearest-neighbor Ising (ANNNI)
model in which there is an isotropic nearest-neighbor ferromagnetic
interaction and a next-nearest-neighbor antiferromagnetic interaction along a
single axis\cite{fisher}.  Even in one dimension, this system exhibits
non-trivial magnetic properties.  For a weak antiferromagnetic interaction,
the ground state is ferromagnetic, while for a strong antiferromagnetic
interaction there is an ``antiphase'' ground state which consists of 2 spins
up, followed by 2 spins down, {\it etc}.  For a specific ratio of these two
interactions, an infinitely degenerate ground state arises in which each spin
domain is of length 2 or greater\cite{sr}.
  
Given the very different natures of these ground states, one might expect
that dynamical behavior is also strongly affected by such competing
interactions.  Our goal is to understand the kinetics of an Ising chain with
nearest-neighbor ferromagnetic interaction $J_1$ and $n^{\rm th}$-neighbor
antiferromagnetic interaction $-J_n$, when the system is also endowed with
single spin-flip Glauber dynamics.  The Hamiltonian of the system is
\begin{equation}
\label{ham}
{\cal H} = -J_1\sum_i s_is_{i+1} + J_n\sum_i s_is_{i+n}.
\end{equation}
For $J_n>J_1/n$, the ground state is a sequence of $n$ consecutive up spins
followed by $n$ down spins, {\it etc}.  Our basic result is that for
$J_n>J_1$ the asymptotic approach of the system to this alternating ground
state can be described in terms of a reactive gas of elementary excitations.
The rate limiting step of the coarsening is governed by the $(n+1)$-particle
annihilation of the dominant excitations, which implies that the system
coarsens in time as $t^{-1/n}$.
  
To provide a context for this work, let us recall the well-known example of
zero-temperature ($T=0$) coarsening in the Ising-Glauber chain with
nearest-neighbor ferromagnetic interactions only\cite{glauber}.  Spin flips
inside ferromagnetic domains are forbidden at $T=0$ because they cost energy,
while spins at the domain interfaces can flip freely, as indicated by
\begin{eqnarray*}
-----&-&++++++\\
  &\Downarrow& \\
-----&+&++++++ \,\, ,
\end{eqnarray*}
since no energy cost is involved.  This spin flip is equivalent to the
hopping of a domain wall ``excitation'' which lies between the neighboring
misaligned spins.  Two domain walls can annihilate when they meet, a
process which leads to a reduction of the energy in the system.  In terms of
the spins, the annihilation is equivalent to a domain which shrinks to zero
size via the process
\begin{eqnarray*}
++++++&-&++++++\\
  &\Downarrow& \\
++++++&+&++++++\,\, .
\end{eqnarray*}
This correspondence between the ferromagnetic Ising-Glauber chain and a gas
of domain walls which undergo nearest-neighbor hopping and single-species
two-particle annihilation provides a simple way to understand the $t^{-1/2}$
coarsening dynamics of the spin system\cite{amar}.  In the following sections
we show that the Ising-Glauber chain with competing interactions can be
understood through a similar picture of reactive excitations. 

\section{SECOND-NEIGHBOR INTERACTION}
  
\subsection{Strong Antiferromagnetism: $J_2>J_1$}

Let us now study the Ising-Glauber chain with competing interactions at
$T=0$.  To be concrete, consider first the case $n=2$, {\it i.e.}, a
near-neighbor ferromagnetic interaction $J_1$ and a second-neighbor
antiferromagnetic interaction $-J_2$.  We focus on the case of strong
antiferromagnetic interaction, $J_2>J_1$, as this is the case which leads to
interesting dynamics.  Starting from an initial ferromagnetically ordered up
state for simplicity, this system evolves to the $\ldots\, ++--++-- \,\ldots
$ ground state by a two-stage process.  First, there is an initial nucleation
of down domains within regions of up spins.  Within a ferromagnetically
ordered region, there is an energy loss $\Delta E=4J_1-4J_2$ when a single
spin flips to create an isolated down spin.  This same energy loss arises for
any nucleation event which occurs within a domain of length $\geq 5$ when the
flipped spin is at least two lattice spacings from both domain walls.  After
this nucleation, the single-spin domain can grow to length 2 with an energy
loss $\Delta E=-4J_2$ as long as the length of the neighboring domain into
which this inflation occurs is greater than 2.  At the end of this nucleation
stage, therefore, the system consists of ordered $\ldots\, ++--++\,\ldots$
regions, as well as domains of size 1, 3, and 4.  Further, domains of size
unity exist only if both neighboring domains are of length 2.

These remaining domains now undergo a sequence of reactions which ultimately
leads to the system reaching the ground state.  To determine this evolution,
first consider an isolated 3-domain within an otherwise stable array to
2-domains.  Since the spin on either edge of the 3-domain can flip without
energy cost, the 3-domain can hop isotropically by two lattice sites, as
indicated by
\begin{eqnarray*}
++--++&-&--++--\\
  &\Downarrow& \\
++--++&+&--++--\,\,.
\end{eqnarray*}
Similarly, an isolated 1-domain within a sea of 2-domains also diffuses
freely, since either neighboring spin of the 1-domain can flip with no energy
cost, as indicated by
\begin{eqnarray*}
--++-&+&+--++--\\
  &\Downarrow& \\
--++-&-&+--++--\,\,.
\end{eqnarray*}
These freely diffusing 1- and 3-domains are the elementary excitations of the
system.
  
To appreciate the consequences of this statement, consider an isolated
4-domain within an otherwise stable array of 2-domains.  Since there is an
energy cost associated with flipping either spin in the interior of a
4-domain, this process does not occur at zero temperature.  If, however,
either spin at the end of the 4-domain flips, for example,
\begin{eqnarray*}
--++---&-&++--++ \\
  &\Downarrow& \\
--++---&+&++--++ \,\, ,
\end{eqnarray*}
then the configuration becomes two adjacent 3-domains within the stable sea
of 2-domains.  Each of these 3 domains can then diffuse freely; the first
step of this process is indicated by
\begin{eqnarray*}
--++&-&--+++--++ \\
  &\Downarrow& \\
--++&+&--+++--++ \,\, .
\end{eqnarray*}
One can thus regard an isolated 4-domain as a  $[3,3]$ resonant state which
is formed whenever two 3-domains collide.  This resonance is short-lived,
however, since its binding energy is zero.
  
Continuing this reasoning, consider the evolution of a $4,3$ pair within a
stable sea of 2's.  Since both the 3 and 4 diffuse freely, these two
excitations could move apart with zero energy cost.  On the other hand, the
3-domain can shrink and the 4-domain can grow to size 5, also with zero
energy cost.  If this occurs, the 5 is unstable to the nucleation event:
$5\to 2+1+2$.  Due to the fact that the 4 can be viewed as a $[3,3]$
resonance, the penultimate stoichiometry of this process is $3+3+3\to 1$,
{\it i.e.}, 3's annihilate through triple collisions.
  
Finally consider how isolated 1's evolve.  Since 1's diffuse freely they
react only upon meeting another 1 or a 3.  In the former collision, a 4 is
formed, 
\begin{eqnarray*}
++--++&-&+--++--\\
  &\Downarrow& \\
++--++&+&+--++-- \,\, ,
\end{eqnarray*}
with an associated energy loss $\Delta E=-4J_1$.  Because the 4 is equivalent
to a pair of 3's, the penultimate stoichiometry of this process is $1+1\to
3+3$.  Finally if a 1 meets a 3, they react to form a stable pair of 2's,
\begin{eqnarray*}
++--+&-&--++--\\
  &\Downarrow& \\
++--+&+&--++-- \,\, .
\end{eqnarray*}
with an associated energy loss $\Delta E=-4J_2$.  This can be viewed as the
two-species annihilation process $1+3\to 0$, since the pair of 2's that are
formed become part of the stable ground state.
  
The underlying stoichiometry of these processes can therefore be summarized by
the three reactions
\begin{equation}
\label{summ}
\begin{array}{lllll}
3+3+3&\to& 1  &\quad \Delta E =&+4J_1 -4J_2, \\
1+1&\to &3+3  &\quad \Delta E =&-4J_1,  \\
1+3&\to &0    &\quad \Delta E =&-4J_2. \\ 
\end{array}
\end{equation}
Since each of these processes leads to a loss of energy, they each occur at
the same rate when $T=0$.  The overall effect of these reactions is that the
density of these elementary excitations ultimately vanish.  Let us now
determine the time dependence for the densities of these excitations in the
mean-field limit.  Using $A$ to denote both domains of length 3 and their
density, and similarly using $B$ for 1's, the rate equations associated with
the reaction scheme (\ref{summ}) are
\begin{eqnarray}
\dot A &=-3&A^3 +2B^2 -AB,\label{rea}\\
\dot B &=  &A^3 -2B^2 -AB.\label{reb}
\end{eqnarray}
A naive qualitative analysis of these equations indicates that $A$'s are
asymptotically dominant.  Thus using $A\gg B$, the above rate equations
simplify to
\begin{eqnarray}
\dot A &\simeq -3&A^3-AB,\label{nrea}\\
\dot B &\simeq   &A^3-AB.\label{nreb}
\end{eqnarray}
Subtracting Eq.~(\ref{nreb}) from (\ref{nrea}) gives $\dot A-\dot B\simeq
-4A^3$.  Given $A\gg B$, we neglect $\dot B$ to give the closed equation
$\dot A\simeq -4A^3$.  Comparing with Eq.~(\ref{nrea}) yields $A^3\simeq AB$.
This implies $B\simeq A^2$, which agrees with $B\ll A$.  The end results of
these considerations are
\begin{equation}
\label{defamp}
A(t)\simeq {1\over \sqrt{8t}} \quad{\rm and}\quad 
B(t)\simeq {1\over 8t},
\end{equation}
which are confirmed by numerical integration of the rate equations.

Let us now adapt the rate equations to the case of one dimension.
Quite generally, we write 
\begin{eqnarray}
\dot A &\simeq -3&R-r,\label{reoneda}\\
\dot B &\simeq  & R-r.\label{reonedb}
\end{eqnarray}
Here $R$ is the rate at which $A$'s disappear due to the triple collisions
and $r$ is the rate at which both $A$'s and $B$'s disappear due to their
mutual annihilation.  These equations are at the same level of approximation
as Eqs.~(\ref{nrea}) and (\ref{nreb}), as we again neglect the effect of
interactions between two $B$'s.  The rate $R$ can be shown to scale as
$A^3/\ln(1/A)$\cite{pk}.  The cubic term is just the mean-field rate of
triple collisions between $A$'s, and the logarithmic correction arises
because one dimension is the critical dimension for the three-particle
reaction-diffusion processes\cite{oh}.  This reaction rate gives the rate
equation $\dot A\sim -A^3/\ln(1/A)$ for three-particle annihilation.  This
gives the asymptotic behavior for the density $A\sim \sqrt{\ln t/t}$, which
agrees with previous numerical and theoretical treatments\cite{dani}.  The
two-species annihilation rate $r$ can be estimated as $r\sim B/\tau$, where
$\tau$ is the reaction time for an $A$ and a $B$ to meet by diffusion.  This
reaction time is proportional to the square of the distance between a $B$ and
its nearest $A$; therefore, $\tau\sim 1/A^2$ which gives $r\sim BA^2$.

Subtracting Eq.~(\ref{reonedb}) from (\ref{reoneda}) gives $\dot A-\dot
B\simeq -4R$.  Since we again anticipate that $A\gg B$, we ignore $\dot B$ to
obtain $\dot A\simeq -4R$.  This relation implies that triple collisions are
the dominant kinetic mechanism for elimination of $A$'s so that this density
decays as in three-particle annihilation.  The factor of 4 in $\dot A\simeq
-4R$ indicates that four $A$'s eventually disappear after a triple collision
-- three particles are eliminated in the process $A+A+A\to B$, and then the
newly-formed $B$ will eliminate another $A$.  To determine $B(t)$, note that
Eq.~(\ref{reoneda}), together with $\dot A\simeq -4R$, imply that $r\simeq
R$; that is, the gain and loss terms in Eq.~(\ref{reonedb}) cancel.  The
relation $r\simeq R$ can be rewritten as $B\sim R/A^2\sim -\dot A/A^2$.  We
therefore conclude that the density of elementary excitations are
\begin{equation}
\label{AB}
A(t)\sim \sqrt{\ln t\over t} \quad{\rm and}\quad 
B(t)\sim {1\over\sqrt{t \ln t}}.
\end{equation}

\begin{figure}
\narrowtext
\epsfxsize=2.5in\epsfysize=2.5in
\hskip 0.3in\epsfbox{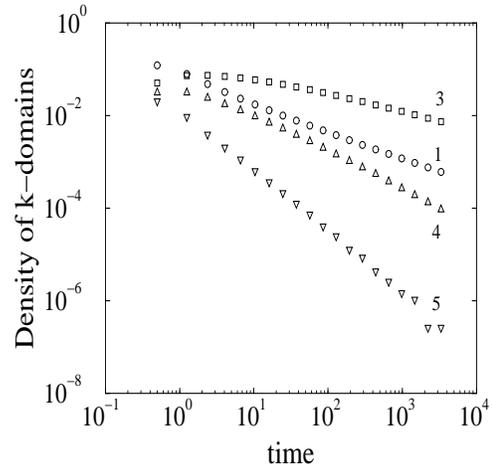}
\vskip 0.15in
\caption{Density of domains of length 1 ($\circ$), 3 ($\Box$), 4
  ($\Delta$) and 5 ($\nabla$).  Simulation results are based on 1000
  realizations of a chain of 20000 sites with $J_2/J_1>1$.
\label{fig1}}
\end{figure}

Monte Carlo simulations of the $T=0$ Ising-Glauber chain yield results which
are qualitatively consistent with these predictions.  As shown in Fig.~1, the
densities of the various elementary excitations decay at different temporal
rates.  Over the last 2 decades of data, linear least-squares fits give 0.56
and 0.40 for the exponent associated with the density of 1's and 3's,
respectively.  However, there is considerable curvature in the data and the
linear fit is not indicative of the asymptotic behavior.  In fact, the data
for the density of 1's is curved upward while that for the density of 3's is
curved downward; these are suggestive of an asymptotic exponent of 1/2.
However, a serious quantitative test of Eqs.~(\ref{AB}) would require
extensive simulation, since the corrections to the local exponents will
vanish only as $1/\ln t$.  We thus interpret the data as being consistent
with (\ref{AB}).  Similar least-squares fits give exponents for the density
of 4's and 5's as 0.84 and 1.42.  In this case, both sets of data exhibit
downward curvature.  These two features are qualitatively consistent with the
hypothesis that 4's and 5's are equivalent to $[3,3]$ and $[3,3,3]$
resonances, respectively, so that their densities should scale as $A^2$ and
$A^3/\ln(1/A)$.

To provide an additional insight into the basic nature of this coarsening
process it is worthwhile to consider the case $J_1=0$, where the chain breaks
up into two independent antiferromagnetic sublattices.  For each sublattice,
the dynamics must coincide with the usual $t^{-1/2}$ Glauber coarsening.  It
is instructive to see how this behavior arises within our picture of
elementary excitations.  The crucial feature for the case $J_1=0$ is that the
process $1+1\to 3+3$ now involves no energy change (see Eq.~(\ref{summ})).
Therefore, the reverse process $3+3\to 1+1$ occurs freely.  The presence of
this additional process now gives the rate equations
\begin{eqnarray}
\dot A &=-3&A^3 -2A^2 +2B^2 - AB,\label{repa}\\
\dot B &=&  A^3 +2A^2 -2B^2 - AB.\label{repb}
\end{eqnarray}
The cubic term turns out to be negligible and the resulting asymptotic
behavior is $A(t)=B(t)\sim 1/t$.  Similarly, for the one-dimensional system,
the density of 1's and 3's become identical and both decay as $t^{-1/2}$.
The presence of this additional reaction process and the corresponding change
in the dominant terms in the rate equations shows clearly that the dynamics
of the Ising-Glauber chain with competing interactions is in a different
universality class than that of the purely antiferromagnetic spin chain.

\subsection{Weak Antiferromagnetism: $J_2\leq J_1$}

To complete the discussion of the Ising-Glauber chain with competing
interactions, let us extend the considerations discussed above to $J_2\leq
J_1$.  For $J_2<J_1$, the basic new dynamical feature is that nucleation of
isolated domains in energetically forbidden at $T=0$.  Further, any 1's in
the initial population of elementary excitations quickly disappear through
the annihilation processes $1+1\to 3+3$ and $1+3\to 0$.  However 3's saturate
at a finite density, as there is no mechanism for their removal.  Resonances
such as $[3,3]$, $[3,3,3]$, {\it etc.}, form freely by collisions between
3's.  However, high-order resonances cannot decay by nucleation but merely
re-fragment into lower-order resonances.

In the case of $J_2<J_1/2$, domain walls become the basic excitations, as in
the ferromagnetic Ising-Glauber chain.  However, domain walls repel since
there is an energy cost $\Delta E=4J_2$ to bring two domain walls closer than
two lattice spacings apart.  Interestingly, for all $J_2<J_1$, the
equilibrium ground state cannot be reached via single spin-flip Glauber
dynamics.  For $J_1/2<J_2<J_1$, the alternating $\ldots ++--++\ldots$ ground
state cannot be attained because there is no mechanism for domains of length
$\geq 3$ to break up; for $J_2<J_1/2$, the ferromagnetic ground state cannot
be attained because of the repulsion of domain walls.

The situations where $J_2=J_1/2$ and $J_2=J_1$ exhibit additional new dynamical
features and each requires a separate treatment.  In the former case, an
arbitrary array of domains which all have lengths $\geq 2$ forms a ground
state\cite{sr}.  The Glauber single spin-flip dynamics now has the following
geometrical manifestations.  First, it is energetically favorable for
isolated 1-domains to disappear by ``anti-nucleation''.  Second, domain walls
can diffuse freely as long as they are more than two lattice spacings apart.
They repel each other with infinite strength at zero temperature, so they
cannot annihilate.  Most interesting is the dynamical behavior of strings of
consecutive antiferromagnetic spins.  In this case, the flipping of a single
spin gives rise to a 3-domain.  This can be viewed as the replacement of the
triplet of antiferromagnetic spins by a ferromagnetic ``trimer''.  This
replacement process of 3 spins by a trimer can continue inside an initially
antiferromagnetic region until it is converted into a ferromagnetic domain.

This dynamics is therefore equivalent to the random sequential adsorption of
trimers which can overlap.  The dynamics of this adsorption process can be
solved according to well-known procedures\cite{evans}.  Let $P_n(t)$ denote
the probability that exactly $n$ consecutive sites are occupied by 1-domains,
that is, an antiferromagnetically ordered string of length $n$.  The
probabilities $P_n(t)$ evolve according to the rate equations
\begin{equation}
\label{rsa}
\dot P_n=-(n+2)P_n+2\sum_{j=n+1}^\infty P_j.
\end{equation}
The loss term arises because the adsorption of a trimer whose center
coincides with any of the $n$ sites of the string or the two sites adjacent
to the string destroys the $n$-string.  Similarly, the gain term arises from
processes in which the adsorption of a trimer onto a larger string leads to
the creation of an $n$-string.

To solve the rate equations, note that the ansatz
\begin{equation}
\label{ansatz}
P_n(t)=F(t) f^{n+2}(t)
\end{equation}
transforms the infinite set of differential equations (\ref{rsa}) into the
pair of equations
\begin{equation}
\label{fF}
\dot f=-f, \quad \dot F={2F f^2\over 1-f}.
\end{equation}
Consider now an initially disordered Ising chain (corresponding to a quench
from $T=\infty$ to $T=0$).  An antiferromagnetic string of length $n$ occurs
with probability $P_n(0)=2^{-n-3}$, which implies
\begin{equation}
\label{initial}
F(0)=f(0)={1\over 2}. 
\end{equation}
Solving Eqs.~(\ref{fF}) subject to (\ref{initial}) yields 
\begin{eqnarray}
f(t)&=&{1\over 2}\,e^{-t}, \label{f}\\
F(t)&=&2\left(1-{1\over 2}\,e^{-t}\right)^2
\exp\left[e^{-t}+{1\over 4}\,e^{-2t}-{5\over 4}\right].\label{F}
\end{eqnarray}
Therefore domain walls inside antiferromagnetic strings quickly disappear and
the total density of domain walls reaches a saturation level.  Since the
subsequent dynamics does not allow for processes which change the number of
domain walls, the system continues to explore a sector of the phase space
that contains all ground states with the same number of domain walls.  The
system falls into its asymptotic sector exponentially in time.  Different
sectors are mutually disconnected, and the sector eventually reached by the
system depends on initial conditions.  The dependence on initial conditions
and exponential relaxation are outcomes of the lack of ergodicity in the
Glauber dynamics of the Ising model with competing interactions when
$J_2=J_1/2$.

Consider finally the marginal case $J_2=J_1$.  The crucial difference with
the case of strong antiferromagnetism, $J_2>J_1$, is that the process
$3+3+3\to 1$ now involves no energy gain and thus the reverse process $1\to
3+3+3$ can also occur freely.  The rate equations (\ref{reoneda}) and
(\ref{reonedb}) are thus modified to
\begin{eqnarray}
\dot A &\simeq  3&B-3R-r,\label{ra}\\
\dot B &\simeq &R-r-B.\label{rb}
\end{eqnarray}
To determine the asymptotic behavior, it proves convenient to transform
these equations to
\begin{eqnarray}
\dot A -\dot B &\simeq& -4(R-B),\label{raa}\\
\dot A+3\dot B &\simeq& -4r.\label{rbb}
\end{eqnarray}
We now ignore $\dot B$ on the left-hand sides.  Combining the estimate
$r\sim BA^2$ with Eq.~(\ref{rbb}) gives $B\sim -\dot A/A^2$. This implies
$B\gg -\dot A$, so that Eq.~(\ref{raa}) now gives $B\simeq R$, or $\dot A\sim
-A^5/\ln(1/A)$.  We therefore conclude that when $J_2=J_1$ the density of
elementary excitations are
\begin{equation}
\label{ab}
A(t)\sim \left({\ln t\over t}\right)^{1/4}\quad {\rm and}\quad  
B(t)\sim \left({1\over t^3\,\ln t}\right)^{1/4}.
\end{equation}
Thus in the marginal case of $J_2=J_1$, the Ising-Glauber chain still
coarsens, but at a much slower rate than in the case of strong
antiferromagnetism, $J_2>J_1$.

\section{THIRD- AND MORE DISTANT-NEIGHBOR INTERACTION}

Our general approach can naturally be applied to longer range
antiferromagnetic interactions.  We first outline basic features of
coarsening for an antiferromagnetic third-neighbor interaction; the behavior
for arbitrary range antiferromagnetic interaction follows inductively.  The
Hamiltonian now is ${\cal H} = -\sum_i( J_1 s_is_{i+1} - J_3 s_is_{i+3})$ and
for sufficiently strong antiferromagnetic interaction the ground state
consists of alternating domains of length 3, $\ldots +++---+++\ldots $.
Starting from the ferromagnetic up state, the evolution to the ground state
again proceeds by a two-stage process when $J_3>J_1$.  While the ground state
occurs when $J_3>J_1/3$, we shall employ the stronger inequality $J_3>J_1$ in
the following to guarantee that this ground state is accessible via single
spin-flip Glauber dynamics.

In the initial nucleation stage, it is energetically favorable for a spin
within an large up domain to flip if this spin is three or more lattice
spacings from any domain boundary.  Thus domains of length $\geq 7$ are
unstable to such nucleation events.  It is also energetically
favorable for this isolated down spin domain to grow to size 3, as long as
the expanding domain wall remains at least 3 lattice spacings from adjacent
domain walls.  At the end of the nucleation stage, therefore, all domains
have length $\leq 6$.  Further, domains of length 1 or 2 must be surrounded
by domains of length 3; otherwise the central domain would expand until its
size reached 3.

A basic observation is that the true elementary excitations are domains of
length 2 and of length 4, as these are the only objects which diffuse freely
within a stable sea of 3's.  All other defects are resonant states of these
two elementary excitations.  To determine the nature of the coarsening, first
consider the resonances of 4-domains.  For example, a 5-domain is formed when
two 4's meet and interact, so that one 4 shrinks to length 3, while the other
grows to length 5.  There is no energy cost associated with this process, so
that a 5-domain can be viewed as a $[4,4]$ resonance.  Similarly, a 6-domain
is a $[5,4]$, or equivalently a $[4,4,4]$ resonance.  Finally, a 7-domain may
be produced by the conversion of $6+4\to 7+3$; thus the 7 is a $[4,4,4,4]$
resonance.  At the center of the 7-domain, a single spin can flip, thereby
nucleating a 1-domain and two surrounding stable 3 domains.  The penultimate
stoichiometry of this process is therefore, $4+4+4+4\to 1$.

Conversely, consider the resonances and interactions associated with
2-domains.  When two 2's, within a stable sea of 3's, meet, a 1-domain is
formed as indicated by
\begin{eqnarray*}
+++--&+&+---+++\\
  &\Downarrow& \\
+++--&-&+---+++ \,\, .
\end{eqnarray*}
Since there is no energy cost associated with this process, a 1-domain is
simply a $[2,2]$ resonance.  When three 2's meet, there are several possible
zero-energy-cost outcomes.  If the interior spin of the outer 2-domain flips,
then the result is $2+2+2\to 1+3+2$.  Since the 1-domain is a $[2,2]$
resonance, this process can be considered as the first step in separating the
three initial domains.  However, if one of the spins in the middle domain
flips, then the outcome is $2+2+2 \to 3+1+2$.  Once a $3,1,2$ state is
reached, it is energetically favorable for the central isolated spin to flip
thus giving $3+1+2 \to 6$, {\it i.e.}, a $[4,4,4]$ resonance.  This last step
is accompanied by the energy loss $-4J_1$.  The outcome of more than three
2's meeting can be obtained by grouping the 2's into triplets and analyzing
the outcome of each triplet in series.  Finally, when a 2 and a 4 meet, it is
energetically favorable for $2\to 3$ and $4\to 3$.  This can be viewed as the
two-species annihilation $2+4\to 0$, since the two 3's formed in the reaction
belong to the stable ground state.

{}From these basic processes, the governing reactions for this system are
\begin{equation}
\begin{array}{llll}
\label{summnew}
4+4+4+4 &\to & 1  & \to 2+2 \\
2+4   &\to & 0  & \\ 
2+2+2  &\to & 6  & \to 4+4+4 \\
\end{array}
\end{equation}
For these reactions, the associated rate equations for the density of 4's
$(A)$ and 2's $(B)$ are
\begin{eqnarray}
\dot A &=-&4A^4 +3B^3 -AB,\label{re3a}\\
\dot B &= &2A^4 -3B^3 -AB.\label{re3b}
\end{eqnarray}
The structure of these equations is similar to the second-neighbor interaction
case.  Following the same reasoning as used previously, we find $A(t)\sim
t^{-1/3}$ and $B(t)\sim t^{-1}$.  Similarly, we may adapt the rate equation
above to describe the system in one dimension by following the approach used
to write Eqs.~(\ref{reoneda}) and (\ref{reonedb}).  This leads to
\begin{eqnarray}
\dot A &\simeq -&4A^4 -A^2B,\label{re3oneda}\\
\dot B &\simeq  &2A^4 -A^2B,\label{re3onedb}
\end{eqnarray}
with the asymptotic behavior $A(t)\sim t^{-1/3}$ and $B(t)\sim t^{-2/3}$.

\begin{figure}
\narrowtext
\epsfxsize=2.5in\epsfysize=2.5in
\hskip 0.3in\epsfbox{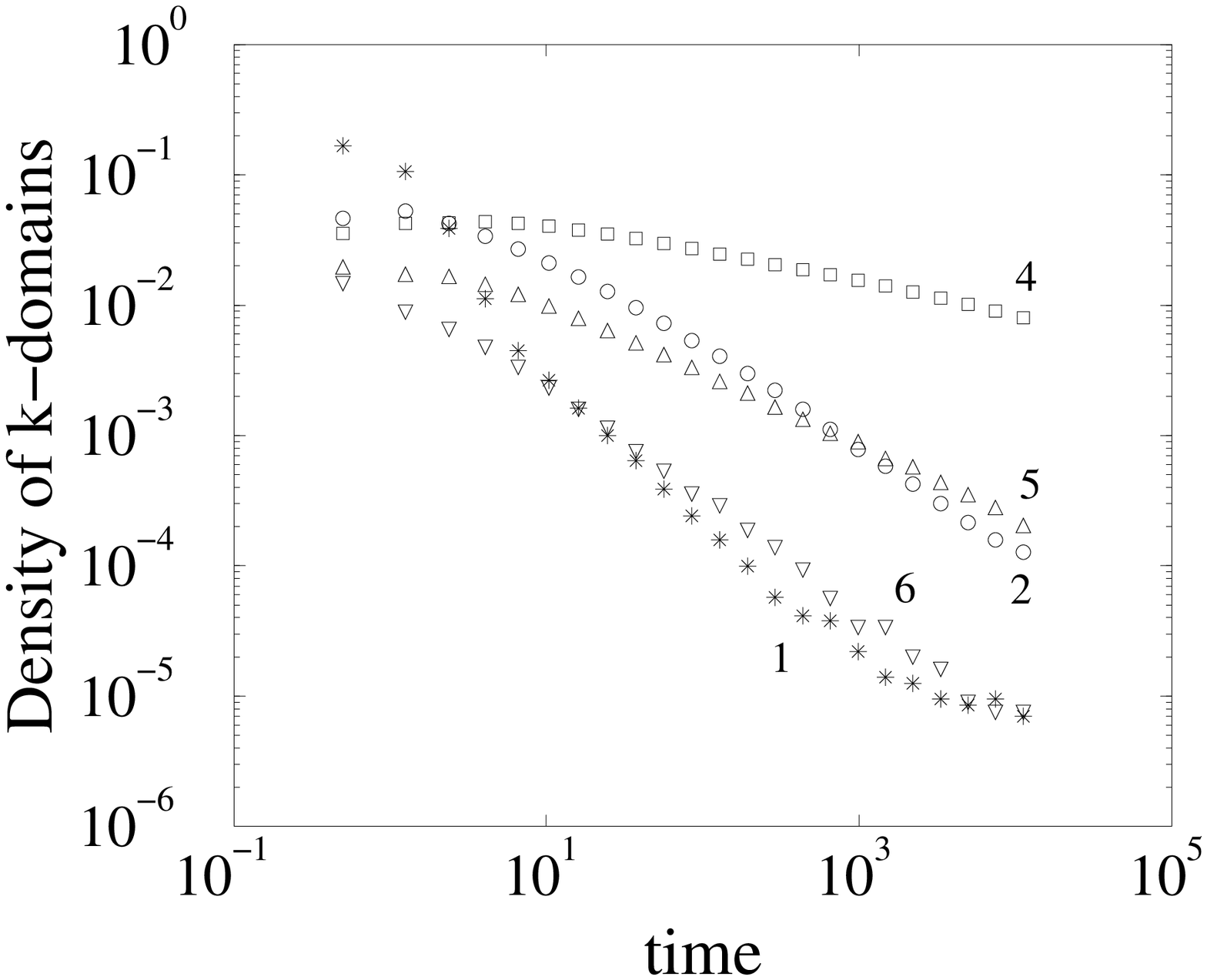}
\vskip 0.15in
\caption{Density of domains of length 1 ($\ast$), 2 ($\circ$), 4 ($\Box$), 5
  ($\Delta$), and 6 ($\nabla$).  Simulation results are based on 500
  realizations of a chain of 20000 sites with $J_3/J_1>1$.
\label{fig2}}
\end{figure}

As in the case of the Ising chain with first- and second-neighbor
interactions, simulations for the densities of the various elementary
excitations indicate that they decay at different temporal rates (Fig.~2).
Over the last 2 decades of data, linear least-squares fits give 0.80 and 0.25
for the exponent associated with the density of 2's and 4's, respectively.
Once again, the sense of the curvature in these two data sets is consistent
with the respective asymptotic exponents of 1/3 and 2/3.  However, even more
so than in Fig.~1 the linear fit is not indicative of asymptotic behavior.
We merely point out that our picture of elementary excitations allows one to
express all densities in terms of the density of 4's.  This predicts that the
density of 1's scales as $A^4$, the densities of 2's and 5's scale as $A^2$,
and the density of 6's scales as $A^3$.  This is only marginally consistent
with the data, a feature which we attribute to slow approach to asymptotic
behavior.

Finally, we may apply a similar geometrical picture of elementary excitations
to treat the general case of the Ising-Glauber chain with competing first-
and $n^{\rm th}$-neighbor interactions when $J_n>J_1$.  In this case, the
basic excitations are domains of $n+1$ spins ($A$) and domains of $n-1$ spins
($B$), each of which diffuses freely within a stable ground state sea of
alternating ferromagnetic strings of length $n$.  Other types of excitations
are resonances of these elementary excitations.  The basic kinetic mechanisms
that govern the coarsening of the spin system are the $(n+1)$-particle
annihilation of contiguous groups of $A$ excitations via $(n+1)A\to 1$, and
the two-species annihilation $A+B\to 0$.  An analysis of the corresponding
rate equations again indicates that these two processes are of the same order
of magnitude.  Consequently, the ground state is approached as $t^{-1/n}$ and
the densities of the basic excitations decay according to $A(t)\sim t^{-1/n}$
and $B(t)\sim t^{-(n-1)/n}$.

\section{CONCLUSIONS}

We have investigated the coarsening kinetics of the Ising chain with single
spin-flip dynamics when a more distant-neighbor antiferromagnetic interaction
$-J_n$ competes with the nearest-neighbor ferromagnetic interaction $J_1$.
For $J_n>J_1$, this competition leads to slower zero-temperature coarsening
compared to the case of nearest-neighbor ferromagnetic interactions only.
Other types of non-universal relaxation phenomena have been reported for
alternating interactions and other modifications of the pure Ising
chain\cite{cornell}.  The case of the competing interaction, however, is
amenable to an intuitively appealing description in terms of elementary
excitations which makes clear the mechanism for the new types of coarsening
kinetics.  It is intriguing that the nature of the elementary excitations and
the spectrum of resonances of the system are not obviously connected with the
microscopic interaction of the spin system.

For general $n^{\rm th}$-neighbor antiferromagnetic interaction,
the elementary excitations are ferromagnetic strings of $n+1$ and $n-1$
spins.  The former interact and disappear through $(n+1)$-particle
single-species annihilation while both excitations mutually annihilate when
they meet.  These two processes are of the same order of magnitude so that
the rate of the overall coarsening process can be viewed as being limited by
$(n+1)$-particle annihilation.  This leads to a coarsening which proceeds as
$t^{-1/n}$.  The marginal case $J_n=J_1$ admits additional microscopic
processes which leads to even slower coarsening.  Finally, it is interesting
to note that Glauber dynamics has only tenuous connection with equilibrium
properties of the system.  That is, the dynamical criticality at $J_n=J_1$ is
disconnected from the corresponding equilibrium behavior, where
ferromagnetism occurs for $J_n<J_1/n$, a ground state of alternating domains
of $n$ ferromagnetic spins occurs for $J_n>J_1/n$, and an infinitely
degenerate ground state consisting of alternating domains of $\geq n$ spins
occurs for $J_n=J_1/n$.

\bigskip\noindent
We gratefully acknowledge NSF grant DMR-9632059 and ARO grant
DAAH04-96-1-0114 for financial support.

\end{multicols} 
\end{document}